# Non-equilibrium Thermal Resistance of Interfaces Between III-V Compounds

Jinchen Han[1], and Sangyeop Lee[1,2*]

[1] *Department of Mechanical Engineering and Materials Science,*

*University of Pittsburgh, Pittsburgh PA 15261*

[2] *Department of Physics and Astronomy,*

*University of Pittsburgh, Pittsburgh PA 15261*

[*]Corresponding email: sylee@pitt.edu

**Abstract**

Interfacial thermal resistance has been often estimated and understood using the Landauer formalism that assumes incident phonons with equilibrium distribution. However, previous studies suggest that phonons are out-of-equilibrium near the interface because of the heat flow through the leads and the scattering of phonons by the interface. In this paper, we report a systematic study on how vibrational spectra mismatch affects the degree of phonon non-equilibrium near an interface, how fast it is relaxed as the phonons diffuse into a lead, and the overall interfacial thermal resistance from the non-equilibrium phonons. Our discussion is based on the solution of the Peierls-Boltzmann transport equation with ab initio inputs for 36 interfaces between semi-infinite group-III (Al, Ga, In) and group-V (P, As, Sb) compound semiconductor leads. The simulation reveals that the non-equilibrium phonons cause significant interfacial thermal resistance for all 36 interfaces, making the overall interfacial thermal resistance two to three times larger than that predicted by the Landauer formalism. We observe a clear trend that the degree of phonon non-equilibrium near an interface and the interfacial thermal resistance from the non-equilibrium phonons increase as the mismatch of the Debye temperature of two lead materials increases. This contrasts with the Landauer formalism's predictions, which show no correlation with the Debye temperature mismatch. The relaxation length of the phonon non-equilibrium varies significantly from 50 nm to 1.5 μm depending on the combination of the lead materials. The relaxation length is proportional to the phonon mean free path of the corresponding lead material but also largely depends on the material in the opposite lead. This suggests the relaxation length cannot be considered an intrinsic property of the corresponding lead material. These findings offer vital insights for understanding non-equilibrium effects on the interfacial thermal transport and optimizing thermal design in devices.

# I. Introduction

The growing demand for highly integrated and miniaturized devices has catapulted interfacial thermal transport to the forefront of thermal management and energy conversion [1-4]. As device interface density increases, the efficiency and reliability of numerous technologies—such as microelectronics, photonics, and thermoelectrics—hinge on interfacial thermal transport [3-6]. Particularly in systems utilizing high thermal conductivity materials like GaN and diamond, the thermal resistance at the interface may surpass that of the bulk material, profoundly impacting the overall thermal transport [7-9]. However, understanding interfacial thermal transport is challenging, given complexities such as atomic structure mismatches and intricate heat carrier interactions at the interfaces [3-5]. Hence, a comprehensive theoretical understanding of interfacial thermal conductance is critical for optimizing thermal design [10].

Landauer formalism stands as a widely-used and effective theoretical tool for assessing interfacial thermal resistance [11]. Typically integrated with transmission function from acoustic mismatch models [12], diffusive mismatch models (DMM) [13,14], or atomistic Green's functions [15,16], the Landauer formalism assumes an equilibrium Bose-Einstein distribution ($f^0$) with constant temperature in leads and a finite temperature drop at the interface to determine the interface thermal conductance [11]. However, two main limitations exist with this assumption: i) the internal phonon scattering results in non-zero temperature gradients in both leads given non-zero heat flux and ii) the interface scattering induces non-equilibrium phonon distribution near the interface.

Recent attempts aimed to address these constraints and more accurately elucidate the interfacial thermal resistance beyond Landauer formalism. Regarding the first limitation, the Landauer formalism was modified such that $f^0$ is substituted with the bulk distribution $f^{bulk}$ representing the phonon distribution in an infinitely large sample under a homogenous temperature gradient [17,18]. This modified Landauer formalism was viewed as superior because the $f^{bulk}$ inherently captures the phonons with out-of-equilibrium due to the temperature gradient and phonon scattering. However, considerable discrepancies exist between predictions from the modified Landauer formalism and outcomes from both non-equilibrium molecular dynamics (NEMD) simulations [19] and experimental data [20-22]. Furthermore, this modified Landauer formalism still neglects the non-equilibrium phonons caused by interface-phonon scattering.

The non-equilibrium phonons near interfaces have recently gained significant attention[23]. At metal-semiconductor junctions, the non-equilibrium effects are dominated by electron-phonon (*e-ph*) scattering because the electron primarily carries heat in metal but cannot be well transmitted to the semiconductor. Past studies have shown the complexity, such as the influences of metal layer thickness, in determining the *e-ph* non-equilibrium [24-26]. However, the impacts of *e-ph* non-equilibrium on interfacial thermal transport are still under debate [25-29]. On the other hand, the interplays between phonon-phonon and interface-phonon scattering cause non-equilibrium among phonons at dielectric interfaces [20,30-32]. So far, various theoretical methods, including the multi-temperature model (MTM) [30], NEMD simulation [31,33], and the Peierls-Boltzmann transport equation (PBE) [20,34-37], have been employed to investigate the phonon non-equilibrium effects. The MTM results showed that each branch of phonons exhibit different temperatures near Si-Ge interface indicating local thermal non-equilibrium near the interface [30]. However, the MTM can underestimate the non-equilibrium effects because of its underlying assumption that all modes in the same branch or group are in equilibrium. The NEMD simulations can naturally involve all phonon anharmonicity and consider complex interfacial lattice structure. Previous studies utilizing modal or spectral analysis emphasize the significant effects of anharmonicity among phonon modes on thermal resistance [31,33]. However, NEMD is usually limited to a length scale that is shorter than the phonon mean free paths in high thermal conductivity materials and hence cannot capture the relaxation of the non-equilibrium phonons. This often means the simulation results can depend on the system size. In contrast, the PBE can describe much longer length scale and provide modal distribution function naturally.

Recently we solved the PBE with the kinetic Monte Carlo (MC) method and quantitatively showed the significant non-equilibrium thermal resistance at an interface shared by semi-infinite Si and Ge leads [20]. The study found that i) the interface scattering leads to significant phonon non-equilibrium near the Si-Ge interface, ii) as phonons diffuse from the interface, the non-equilibrium distribution is relaxed to the bulk distribution due to internal phonon scattering, and iii) the entropy generation and consequent thermal resistance during this relaxation process can be substantial, given the large mismatch between the non-equilibrium distribution near the interface and $f^{\text{bulk}}$ far away from the interface. Consequently, the relaxation of the non-equilibrium phonon distribution results in a higher thermal resistance than the interface scattering at the Si-Ge interface. Overall, when considering both mechanisms together – interface scattering and the relaxation of non-equilibrium phonons – the total interfacial thermal resistance of the Si-Ge interface considerably exceeds Landauer formalism's predictions. This observation naturally leads to the questions of whether the significant non-equilibrium resistance is common among many other interfaces. Also, the major factors that determine the degree of phonon non-equilibrium and its decay rate need to be identified.

In this study, we apply the established kinetic MC method to solve the steady-state PBE for 36 interfaces consisting of semi-infinite III-V leads. With the varying atomic mass and interatomic force constants of III-V compounds, their 36 interfaces provide a systematic way of changing the phonon properties and studying their effects on the non-equilibrium phenomena for interfacial thermal transport. Moreover, the III-V compounds and their interfaces are of practical importance for electronic and photonic devices [38-40]. A deeper comprehension of their interfacial thermal transport can guide better device-level design for superior thermal management [40,41].

## II. Method

We solve the PBE for interfaces shared by semi-infinite III-V compound semiconductor leads at 300K. We use the variance reduced kinetic MC method to solve the PBE as elaborated in prior study [20,42]. The method only accounts for the deviational energy distribution from the global equilibrium (GEQ) distribution to reduce the variance. While the details of MC simulation for interfacial thermal transport can be found in our previous study [20], we will briefly discuss the method here. The deviational energy form of the steady-state PBE under the relaxation time approximation (RTA) can be written as

$$v_{x,i} \frac{de_i^d(x)}{dx} = -\frac{e_i^d(x) - e_i^{d,loc}(x)}{\tau_i}, \qquad (1)$$

where $v_{x,i}$ and $\tau_i$ denote the phonon group velocity along x-direction and lifetime of mode $i$, respectively. The $e_i^d(x)$ is the deviation of energy distribution from global equilibrium, which is $\hbar\omega_i \left[ f_i(x) - f_i^{GEQ} \right]$. Here, $\omega_i$ and $f_i$ are phonon frequency and distribution of mode $i$, $\hbar$ is the reduced Planck constant. The $e_i^{d,loc}$ is the deviation of local equilibrium energy distribution ($e_i^{loc}$) from the global equilibrium energy distribution ($e_i^{GEQ}$), i.e., $e_i^{d,loc} = e_i^{loc}(x) - e_i^{GEQ}$.

Our simulation assumes an interface at $x = 0$, shared by two semi-infinite leads. In practice, these semi-infinite leads are approximated as finite-length leads with length of $L_1$ and $L_2$. Here, 1 and 2 denote the left-side and the right-side lead, respectively. These lengths uphold $f^{bulk}$ with the assigned local temperatures $T_H$ and $T_C$ at the hot ($x = -L_1$) and cold boundary ($x = L_2$), respectively. As such, the sampling particles emitted from $x = -L_1$ and $x = L_2$ follow the boundary conditions:

$$e_i^d \Big|_{x=-L_1} = \hbar\omega_i \left[ f_i^0(T_H) - v_{x,i}\tau_i \frac{\partial f_i^0}{\partial T} \frac{dT}{dx} - f_i^{GEQ} \right] \text{ for modes } i \text{ with } v_{x,i} > 0, \qquad (2)$$

$$e_j^d \Big|_{x=L_2} = \hbar\omega_j \left[ f_j^0(T_C) - v_{x,j}\tau_j \frac{\partial f_j^0}{\partial T} \frac{dT}{dx} - f_j^{GEQ} \right] \text{ for modes } j \text{ with } v_{x,j} < 0, \qquad (3)$$

where $f_i^0(T_H)$ and $f_j^0(T_C)$ denote the Bose-Einstein distribution at $T_H$ and $T_C$ of two materials. The $i$ and $j$ denote phonon modes of left lead 1 and right lead 2, respectively.

For the finite leads to represent the semi-infinite leads, the following two conditions should be satisfied. First, the length of the finite leads must ensure the non-equilibrium distribution near the interface is fully relaxed to $f^{bulk}$ within the finite lead. Each lead length for all cases is selected to be $200\Lambda_{avg}$ where $\Lambda_{avg}$ is the mode-averaged mean free path (MFP) of phonons weighted by the modal specific heat. The $\Lambda_{\text{avg}}$ and lead lengths for each III-V semiconductors are summarized in Supplemental Material Table I [43]. The second condition is about the temperature gradient at the boundary in Eq. (2) and (3). While the temperatures at the two boundaries, $T_{\text{H}}$ and $T_{\text{C}}$, are assumed, the temperature gradients in Eq. (2) and (3) are unknown and need to be found from the MC simulation. Thus, the PBE was solved iteratively with varying temperature gradient at the two boundaries until the actual temperature gradient matches with the one assumed for the boundary conditions.

All inputs for the PBE solver, including harmonic phonon properties and three-phonon scattering rates, are calculated from first principles via the VASP [44-47], Phonopy [48] and ShengBTE [49] packages. We employed the local density approximation (LDA) with the projector augmented wave method (PAW) as the exchange correlation functional [50,51]. The lattice structure is relaxed with a primitive cell until the energy change between two electronic steps is less than $10^{-7}$eV. The second and third order interatomic force constants (IFCs) are then obtained using finite-difference methods with a 4×4×4 supercell. The cutoff energy is 600eV for both relaxation and the IFCs calculation. The reciprocal space for electrons is sampled with a 3×3×3 mesh for the 4×4×4 supercell. For the third order IFCs, we set the interaction cutoff range to include up to the third nearest neighbors following the previous literature [52]. The calculated lattice constants and thermal conductivities are listed in Supplemental Material Table I [43]. The reciprocal space for phonons is sampled with 15×15×15 mesh. The phonon transmission function across interfaces was calculated using the DMM and *ab initio* phonon dispersion.

Thermophysical properties such as the local temperature $T_{\text{loc}}$, heat flux $q''$, local entropy generation rate $\dot{S}$, and thermal resistance $R$ can be calculated by post-processing the local phonon distribution from the MC solution of PBE. Each lead is divided into 10 equal-sized control volumes that have a spatially averaged distribution function. The $T_{\text{loc}}$ of each control volume was found as $T_{\text{loc}} = \left(NV_{\text{uc}}\sum_i C_{V,i}/\tau_i\right)^{-1}\sum_i e_i^{\text{d}}/\tau_i$ [53,54], where $N$ is the number of wavevectors in the reciprocal space, $V_{\text{uc}}$ is the volume of unit cell, and $C_{V,i}$ is the volumetric specific heat of mode $i$. The $q''$ is calculated as $q'' = \left(NV_{\text{uc}}\right)^{-1}\sum_i v_{x,i}e_i^{\text{d}}$. The total thermal resistance of the computational domain is $R_{\text{tot}} = (T_{\text{H}} - T_{\text{C}})/q''$. This can be divided into two parts based on their mechanisms: i) intrinsic thermal resistance $R_{\text{bulk}}$ resulting from the bulk thermal resistivity, and ii) the interfacial resistance $R_{\text{int}}$ that is caused by the interface. The $R_{\text{bulk}}$ is simply found as $L_1^{-1}\kappa_1^{-1} + L_2^{-1}\kappa_2^{-1}$, where $\kappa$ is the thermal conductivity reflecting three-phonon scattering in bulk materials. The $R_{\text{int}}$ is obtained by subtracting $R_{\text{bulk}}$ from $R_{\text{tot}}$. This resistance can be further divided into two parts with different mechanisms: i) the resistance due to the relaxation of non-equilibrium phonons to bulk distribution by three-phonon scattering ($R_{\text{neq}}$), and ii) the resistance directly caused by the interface scattering ($R_{int}^0$). The resistance of the whole domain including leads 1 and 2 now becomes $R_{tot} = R_{bulk,1} + R_{neq,1} + R_{int}^0 + R_{neq,2} + R_{bulk,2}$.

In practice, the $R_{neq}$ can be evaluated by calculating the local entropy generation rate due to the three-phonon scattering ($\dot{S}$). The $\dot{S}$ under RTA is [20,55]

$$\dot{S} = \frac{1}{NV_{uc}T_{loc}^2} \sum_i \frac{\left(e_i^d - e_i^{d,loc}\right)^2}{\left(de_i^{loc}/dT\right)\tau_i}. \tag{4}$$

The rate of entropy generation in bulk material without interface ($\dot{S}_{bulk}$) can be found using $e_i^d = e_i^{bulk} - e_i^{GEQ}$ in Eq. (4). Thus, $\dot{S} - \dot{S}_{bulk}$ is the additional entropy generation rate due to the relaxation of non-equilibrium phonons to bulk distribution. The local resistivity from the relaxation process of non-equilibrium phonons ($R'_{neq}$) can be then calculated as $\left(T_{loc}/q''\right)^2 \left(\dot{S} - \dot{S}_{bulk}\right)$. The thermal resistance from non-equilibrium phonons $R_{neq,X}$ of material X is simply obtained by spatially integrating $R'_{neq,X}$. Finally, the $R_{int}^0$ can be found by subtracting $R_{bulk}$ and $R_{neq,A+B}$ (the sum of $R_{neq}$ of material A and B) from $R_{tot}$.

We considered 8 zincblende III-V compounds that are AlP, GaP, InP, AlAs, GaAs, InAs, GaSb, and InSb. We excluded AlSb since the RTA is known to fail for this material [52]. These 8 materials can form 28 heterostructure interfaces and 8 homojunction interfaces.

## III. Results & Discussion

Figure 1(a) illustrates a comparison between the interfacial thermal resistance obtained from the MC simulation and the Landauer formalism combined with DMM, denoted as $R_{\text{int}}$ and $R_{\text{L}}$, respectively. Notably, $R_{\text{int}}$ values are larger than those of $R_{\text{L}}$ by approximately threefold across all 36 interfaces. Fig. 1(b-c) breaks down $R_{\text{int}}$ into $R_{\text{int}}^0$ and $R_{\text{neq,A+B}}$, revealing that the considerable non-equilibrium effects contribute to the difference between $R_{\text{int}}$ and $R_{\text{L}}$.

In Fig. 1(b), $R_{\text{int}}^0$ differs substantially from $R_{\text{L}}$, particularly for interfaces of X-InAs and X-InSb (pink and red points). The $R_{\text{L}}$ can be expressed as

$$R_{\text{L}} = \lim_{\Delta T \to 0} \Delta T \left( \frac{1}{N_1 V_{\text{uc},1}} \sum_{i, v_{\text{x},i} > 0} v_{\text{x},i} \, e_i^{\text{d,loc}} \Big|_{x=0^-} t_{i,1 \to 2} - \frac{1}{N_2 V_{\text{uc},2}} \sum_{j, v_{\text{x},j} < 0} v_{\text{x},j} \, e_j^{\text{d,loc}} \Big|_{x=0^+} t_{j,2 \to 1} \right)^{-1}, \quad (5)$$

where $\Delta T$ is the temperature difference across the interface, $t_{i,1 \to 2}$ and $t_{j,2 \to 1}$ are the transmission function of mode $i$ from lead 1 to 2 and mode $j$ from lead 2 to 1, respectively. Using the fact that the net heat flux is zero when two leads are at the same temperature, Eq. (5) is usually simplified to $R_{\text{L}} = N_1 V_{\text{uc},1} \left[ \sum_{i, v_{\text{x},i} > 0} v_{\text{x},i} (de_i^0/dT) t_{i,1 \to 2} \right]^{-1}$. Similar to Eq. (5), $R_{\text{int}}^0$ in the MC simulation can be expressed as

$$R_{\text{int}}^0 = \lim_{\Delta T \to 0} \Delta T \left( \frac{1}{N_1 V_{\text{uc},1}} \sum_{i, v_{\text{x},i} > 0} v_{\text{x},i} \, e_i^{\text{d}} \Big|_{x=0^-} t_{i,1 \to 2} - \frac{1}{N_2 V_{\text{uc},2}} \sum_{j, v_{\text{x},j} < 0} v_{\text{x},j} \, e_j^{\text{d}} \Big|_{x=0^+} t_{j,2 \to 1} \right)^{-1}. \quad (6)$$

Eq. (6) can be further simplified using $e_i^{\text{d}} = e_i^{\text{d,loc}} + \Delta e_i^{\text{d}}$ and $e_j^{\text{d}} = e_j^{\text{d,loc}} + \Delta e_j^{\text{d}}$ as

$$R_{\text{int}}^0 = \left[ R_{\text{L}}^{-1} + \lim_{\Delta T \to 0} \Delta T \left( \frac{1}{N_1 V_{\text{uc},1}} \sum_{i, v_{\text{x},i} > 0} v_{\text{x},i} \Delta e_i^{\text{d}} \Big|_{x=0^-} t_{i,1 \to 2} - \frac{1}{N_2 V_{\text{uc},2}} \sum_{j, v_{\text{x},j} < 0} v_{\text{x},j} \Delta e_j^{\text{d}} \Big|_{x=0^+} t_{j,2 \to 1} \right)^{-1} \right]^{-1}, \quad (7)$$

where $\Delta e_i^{\text{d}}$ and $\Delta e_j^{\text{d}}$ are the non-equilibrium portions of the distribution functions. Eq. (7) shows that $R_{\text{int}}^0$ depends not only the equilibrium distribution change resulting from temperature drop across the interface ($R_{\text{L}}^{-1}$) but also on the non-equilibrium distribution $\Delta e_i^{\text{d}}$ and $\Delta e_j^{\text{d}}$ at $x=0$.

The Landauer formalism ignores the latter in Eq. (7) as it assumes the equilibrium distributions in two leads. Our previous results of Si-Ge interface indicate a negligible difference between $R_{\text{int}}^0$ and $R_{\text{L}}$, as plotted with a triangle in Fig. 1(b) [20]. This suggests that, for the Si-Ge interface, the equilibrium component primarily drives the change in phonon distribution across the interface. However, when considering 36 interfaces of various III-V compounds in this work, most of the interfaces exhibit significant differences between $R_{int}^0$ and $R_L$. This indicates that changes in the non-equilibrium component across the interface can be as significant as the equilibrium component change. Thus, $R_{int}^0$ depends on both the transmission function and the on-site phonon non-equilibrium distribution.

Another reason for $R_{int}$ being larger than $R_L$ is $R_{neq,A+B}$. As shown in Fig. 1(c), the $R_{neq,A+B}$ is comparable to $R_L$ for all interfaces we studied, similar to the previously studied Si-Ge case [20]. The data from the 36 interfaces in Fig. 1(c) underscore the importance of non-equilibrium effect in predicting and understanding the interfacial thermal resistance in general. The $R_{neq,A+B}$ and $R_L$ have a positive correlation, indicating that the interfaces with a larger $R_L$ exhibit more pronounced non-equilibrium effect.

We confirm that the non-equilibrium effects play a significant role in interfacial thermal resistance. However, calculating $R_{neq,A+B}$ can be complex since it necessitates solving the PBE in both real and reciprocal spaces. Thus, a simple indicator for $R_{neq,A+B}$, if one exists, would be helpful to roughly estimate the non-equilibrium effect. The Debye temperature ratio of the constituent materials can be considered as a viable indicator because it approximately captures the mismatch of acoustic vibrational spectra between two materials. The Debye temperature can serve as a rough measure of acoustic phonons properties to estimate thermal conductivity [56] and the interfacial thermal conductance within the elastic interfacial scattering scheme [3]. In this study, we calculate the Debye temperature using elastic constants from first-principles [57,58], and the Debye temperature values for all studied materials are available in the Supplementary Material Table I [43]. We then define the Debye temperature ratio $\Theta_A/\Theta_B$ as the higher Debye temperature divided by the lower one for a given interface. This ensures that $\Theta_A/\Theta_B$ is always greater than or equal to one. The subscripts A and B denote the constituent materials with higher and lower Debye temperatures, respectively.

Figure 2 shows $\Theta_A/\Theta_B$ to be an effective indicator for estimating $R_{neq,A+B}$, while not for $R_L$ and $R_{int}^0$. It appears in Fig. 2(a-b) that neither $R_L$ nor $R_{int}^0$ has a clear correlation with $\Theta_A/\Theta_B$. This absence of correlation arises since optical modes carry significant heat in the Landauer formalism with DMM. For instance, the optical modes of InSb contribute 13~44% to the total heat flux in X-InSb interfaces. Therefore, the Debye temperature ratio, emphasizing only the acoustic mode mismatch, does not clearly correlate with resistance values derived from Landauer formalism. However, Fig. 2(c) shows a clear correlation between $\Theta_A/\Theta_B$ and $R_{neq,A+B}$, indicating that the mismatch of acoustic vibrational spectra is closely related to the non-equilibrium resistance. In contrast to the Landauer formalism's case, the optical phonons carry minimal heat, attributed to its short mean free paths when considering the internal phonon scattering. For instance, the optical modes of InSb contribute less than 5% to the heat flux in X-InSb interfaces as will be shown later in Fig. 5(d).

In Fig. 2(c), the $R_{\text{neq,A+B}}$ of all 36 interfaces are clustered into two distinct groups (red/pink and other colors). We propose a dimensionless resistance, $R^*_{\text{neq,A+B}}$, defined as $R_{\text{neq,A}}/R_{\text{b,A}} + R_{\text{neq,B}}/R_{\text{b,B}}$. Here, $R_{\text{b,X}}$ represents the ballistic resistance of material X, expressed as $R_{\text{b,X}} = NV_{\text{uc}} \left[ \sum_i |v_{x,i}| (de_i^0/dT) \right]^{-1}$, to eliminate the effects of different group velocity and specific heat of lead materials. Fig. 2(d) compares the $R^*_{\text{neq,A+B}}$ with the Debye temperature ratio. In contrast to $R_{\text{neq,A+B}}$, the dimensionless non-equilibrium resistance does not exhibit two distinct groupings; instead, all 36 interfaces cluster into one group, demonstrating a clear positive correlation with the Debye temperature ratio.

The dimensionless non-equilibrium resistance contains two important parameters that decide the $R_{\text{neq,X}}$: the degree of non-equilibrium distribution at the interface and its duration in space as phonons diffuse into a lead. Using the right lead as an example, the local non-equilibrium resistivity at the interface can be written as

$$R'_{\text{neq}}\Big|_{x=0^+} = \left(\frac{T_{\text{loc}}}{q''}\right)^2 \frac{k_B}{NV_{\text{uc}}} \sum_i \frac{(f_i - f_i^0)^2 - (f_i^{\text{bulk}} - f_i^0)^2}{f_i^0(f_i^0 + 1)\tau_i}\Bigg|_{x=0^+}. \quad (8)$$

By introducing a function $\alpha_i$, defined as $(f_i - f_i^0)^2\big|_{x=0^+} = \alpha_i^2 (f_i^{\text{bulk}} - f_i^0)^2\big|_{x=0^+}$, we can simplify Eq. (8) to

$$R'_{\text{neq}}\Big|_{x=0^+} = \left(\frac{T_{\text{loc}}}{q''}\right)^2 \frac{k_B}{NV_{\text{uc}}} \sum_i \frac{(\alpha_i^2 - 1)(f_i^{\text{bulk}} - f_i^0)^2}{f_i^0(f_i^0 + 1)\tau_i}\Bigg|_{x=0^+}. \quad (9)$$

The $\alpha_i$ represents the degree of non-equilibrium distribution at the interface $x=0^+$. We then assume exponential decay of the non-equilibrium resistivity in space with a decay parameter $\beta_i$. The local non-equilibrium resistivity at any position $x$ can be written as

$$R'_{\text{neq}}\Big|_x = \left(\frac{T_{\text{loc}}}{q''}\right)^2 \frac{k_B}{NV_{\text{uc}}} \sum_i \frac{(\alpha_i^2 - 1)(f_i^{\text{bulk}} - f_i^0)^2}{f_i^0(f_i^0 + 1)\tau_i}\Bigg|_{x=0^+} \exp\left(-\frac{x}{\beta_i |v_{x,i}|\tau_i}\right). \quad (10)$$

The parameter $\beta_i$ signifies the decay rate of a specific modal non-equilibrium resistivity $R'_{\text{neq},i}\big|_{x=0^+}$ in space relative to its mean free path $|v_{x,i}|\tau_i$. The assumption of exponential decay is well justified

by the MC simulation results of Si-Ge interfaces in our previous work [20] and current work for III-V interfaces shown in Fig. 3(b-c). The non-equilibrium resistance of the right lead is written as

$$R_{neq} = \int_0^\infty R'_{neq}\big|_x dx = \sum_i R'_{neq,i}\big|_{x=0^+} \beta_i |v_{x,i}| \tau_i. \quad (11)$$

We replace the heat flux $q''|_{x=0^+}$ with the heat flux of bulk distribution $q''|_{x\to\infty} = (NV_{uc})^{-1} \sum_i \hbar\omega_i v_{x,i}^2 \tau_i (df_i^0/dT)(-dT/dx)$. Then, after normalizing $R_{neq}$ with $R_b$ as aforementioned, the dimensionless non-equilibrium resistance $R^*_{neq}$ can be written as

$$R^*_{neq} = \frac{\sum_i (\alpha_i^2 - 1)\beta_i \hbar\omega_i v_{x,i}^2 |v_{x,i}| \tau_i^2 (df_i^0/dT)}{\left[\sum_i \hbar\omega_i v_{x,i}^2 \tau_i (df_i^0/dT)\right]^2 \left[\sum_i \hbar\omega_i |v_{x,i}|(df_i^0/dT)\right]^{-1}}. \quad (12)$$

Thus, the dimensionless non-equilibrium resistance is mode average of $(\alpha_i^2 - 1)\beta_i$. Thus, $R^*_{neq,A+B}$ can be written as $\langle(\alpha_i^2 - 1)\beta_i\rangle + \langle(\alpha_j^2 - 1)\beta_j\rangle$ where the angle bracket denotes mode average. Figure 2(d) clearly shows that the combined average of $\alpha$ and $\beta$, represented by $R^*_{neq,A+B}$ increases as the two leads have more significant mismatch of the acoustic vibrational spectra.

We further focus on the degree of non-equilibrium at the interface, $\langle\alpha^2\rangle$, which can be estimated by normalizing the non-equilibrium resistivity at the interface by the bulk resistivity of the lead. Using Eq. (9) and $R'_{bulk} = NV_{uc}\left[\sum_i \hbar\omega_i v_{x,i}^2 \tau_i (df_i^0/dT)\right]^{-1}$, the dimensionless resistivity at $x = 0^+$ is written as

$$\frac{R'_{neq}\big|_{x=0^+}}{R'_{bulk}} = \frac{\sum_i (\alpha_i^2 - 1)\hbar\omega_i v_{x,i}^2 \tau_i (df_i^0/dT)}{\sum_i \hbar\omega_i v_{x,i}^2 \tau_i (df_i^0/dT)}, \quad (13)$$

which represents the mode average of $\alpha_i^2 - 1$. Thus, $\langle\alpha_i^2\rangle$ can be calculated as $R'_{neq}\big|_{x=0^+}/R'_{bulk} + 1$. In this work, the $R'_{neq}\big|_{x=0^+}$ or $R'_{neq}\big|_{x=0^-}$ is calculated by fitting $R'_{neq}$ of control volumes with an exponential function.

Figure 2(e) plots the degree of non-equilibrium distribution at the interface $\langle\alpha_i^2\rangle + \langle\alpha_j^2\rangle$ versus the Debye temperature ratio. We observe a strong positive correlation between $\langle\alpha_i^2\rangle + \langle\alpha_j^2\rangle$ and the $\Theta_A/\Theta_B$, suggesting that the acoustic vibrational mismatch is a deterministic factor for the phonon non-equilibrium at an interface. Figure 2(e) also reveals that the phonon distribution at interface is highly non-equilibrium in all interfaces. The values of $\langle\alpha_i^2\rangle + \langle\alpha_j^2\rangle$ range from 2.5 to

4 meaning that the deviation of phonon distribution from equilibrium at the interface up to twice larger than the deviation in bulk case.

Comparing $R_{neq,A}$ and $R_{neq,B}$ across interfaces, we observe that $R_{neq}$ is always more pronounced in the constituent material B with the lower Debye temperature. In Fig. 3(a), the ratios of the non-equilibrium resistances, $R_{neq,A}/R_{neq,B}$, are plotted against $\Theta_A/\Theta_B$. The $R_{neq,A}/R_{neq,B}$ is less than one for all interfaces, indicating that the material A with the higher Debye temperature has a lower non-equilibrium resistance. Moreover, the $R_{neq,A}/R_{neq,B}$ has a positive correlation with $\Theta_A/\Theta_B$. The relationship can be understood by considering the fact that the spectral heat flux is the same across the interface, i.e., $\sum_i v_{x,i} e_i^d \delta(\omega_i - \omega) = \sum_j v_{x,j} e_j^d \delta(\omega_j - \omega)$ when elastic interface scattering is assumed [20]. The material B with lower $v_x$ requires larger deviational distribution $e^d$ to have the same spectral heat flux, thereby generating more substantial entropy based on the Eq. (6). For instance, Fig. 3(b-c) showcase the resistivity $R'$ (black dots) and its exponential fit (black line) across AlAs-InSb and InAs-InSb interfaces. The area between the black and red horizontal lines represents $R_{neq}$. In both heterostructures, the $R_{neq,InSb}$ is significantly larger than both $R_{neq,AlAs}$ and $R_{neq,InAs}$ since the group velocity of InSb is smaller than that of AlAs and InAs.

We now turn our attention to the decay length of the non-equilibrium resistivity. The $R'_{neq}$ can be fitted well with an exponential function, $R'_{neq,i}\big|_{x=0^+} \exp(-x/\lambda)$, as depicted in Fig. 3(b-c). Here we define the fitting coefficient $\lambda$ as a relaxation length. Figure 4 plots the $\lambda$ versus $\Lambda_{avg}$ of the lead. The error bar shows the 95% confidence bounds of the fitting coefficient $\lambda$. Generally, the $\lambda$ is proportional to the $\Lambda_{avg}$ of the corresponding lead material. It is worth noting that the $\lambda$ also depends on the material in the opposite lead. For example, the $\lambda$ in AlP of X-AlP varies by a factor of 10 depending on the material X in the opposite lead. For all interfacial structures, the $\lambda$ varies at least by a factor of 2 to 3. Therefore, the relaxation length $\lambda$ in a lead is not an intrinsic property of the corresponding lead material, but rather depends on the combination of the two materials that constitute the interface.

Like the degree of non-equilibrium distribution represented by $\langle \alpha_i^2 \rangle$, the relaxation length $\lambda$ is also influenced by the Debye temperature ratio. Figure 5 details the relaxation lengths of InSb ($\lambda_{InSb}$) for all X-InSb interfaces. We chose InSb as the lead of interest since it has the lowest Debye temperature among the III-V compounds we studied and thus is expected to exhibit the largest non-equilibrium effects as observed in Fig. 3. Figure 5(a) shows the exponential fitting of the non-equilibrium resistivity $R'_{neq,InSb}$ in all X-InSb interfaces. The non-equilibrium resistivity at the interface, $R'_{neq}\big|_{x=0^+}$, has the largest value in the AlP-InSb heterostructure, gradually decreases as the Debye temperature difference between X and InSb decreases, and has the lowest value in the InSb-InSb interface. In Fig. 5(b), we normalize the $R'_{neq,InSb}$ by $R'_{neq}|_{x=0^+}$ to compare $\lambda_{InSb}$. The $\lambda_{InSb}$ has the opposite trend as the $R'_{neq}|_{x=0^+}$; it is the largest in the InSb-InSb and the smallest in the AlP-InSb. Figure 5(c) summarizes both $R'_{neq}|_{x=0^+}$ and $\lambda_{InSb}$ shown in Fig. 5(a-b) as functions of $\Theta_X/\Theta_{InSb}$. The $R'_{neq}|_{x=0^+}$ increases with $\Theta_X/\Theta_{InSb}$ indicating that a larger acoustic mismatch results in a more out-of-equilibrium phonon distribution at the interface. The $\lambda_{InSb}$ decreases with $\Theta_X/\Theta_{InSb}$ in X-InSb interface structures.

The dependence of relaxation length on the Debye temperature can be understood by measuring the spectral contribution to the thermal transport. Figure 5(d) showcases the spectral heat flux in the control volume centered at $x=0.12\mu m$ which is adjacent to the interface ($x=0\sim0.24\mu m$) in InSb lead. We divide the entire phonon spectrum of InSb into 5 frequency bins and plot the contribution from each bin to the heat flux. The spectral heat flux in Fig. 5(d) is similar to the bulk case when $\Theta_X/\Theta_{InSb}$ is close to 1. However, with an increase in $\Theta_X/\Theta_{InSb}$, contributions from low frequencies diminish, while those from high frequencies grow. This is because the Debye temperature ratio greatly influences the spectral overlap of phonon density-of-states (DOS). In turn, this overlap dictates the spectral transmission function in DMM. Figure 5(e) shows that GaSb-InSb, with its relatively small $\Theta_{GaSb}/\Theta_{InSb}$, exhibits a moderate phonon DOS overlap in the first two frequency bins. In contrast, Fig. 5(f) shows that AlP-InSb has almost negligible phonon DOS overlap in the first two frequency bins. Most of the phonon DOS overlap is concentrated in the higher frequency ranges (bins 3 to 5). As illustrated in Fig. 5(d), this makes the heat flux from high frequency ranges more significant. The MFP of acoustic phonon modes is strongly frequency-dependent in acoustic branches, typically following $\omega^l$ where $l$ ranges from 2 to 3. Therefore, the overall decay length of $R'_{neq}$ depends on the spectral heat flux contribution. When the Debye temperature ratio increases, high-frequency phonons take on a greater role in heat flux. This leads to faster relaxation processes due to the stronger scattering rates associated with these high-frequency phonons, resulting in the shorter decay length of the $R'_{neq}$.

## IV. Conclusion

In this work, we comprehensively analyzed the thermal resistance across 36 III-V interfaces by solving the Peierls-Boltzmann transport equation with ab initio phonon dispersion and three-phonon scattering rates. The DMM is assumed for the phonon transmission across the interface. Our simulations revealed significant non-equilibrium effects for the interfacial thermal resistance. For all 36 interfaces, the overall interfacial thermal resistance considering non-equilibrium phonons is two to three times larger than the interfacial thermal resistance by the Landauer formalism which assumes equilibrium distribution of phonons. We also found the non-equilibrium effect is always pronounced in a lead with lower Debye temperature between two leads constituting the interface. Using the dimensionless form of non-equilibrium resistance, we estimated the degree of phonon non-equilibrium near interface. All 36 interfaces show a clear correlation between the degree of phonon non-equilibrium near interface and the Debye temperature mismatch; it indicates that the acoustic vibrational spectra mismatch is the main factor that determines the non-equilibrium phonons and hence interfacial thermal resistance. It contrasts with the Landauer formalism from which interfacial resistance show no correlation with the Debye temperature mismatch. The relaxation lengths of the non-equilibrium phonons are discussed to identify the extent of space where the non-equilibrium effect is significant. Depending on the interface, the relaxation length varies between 50nm and 1.5μm. In general, while the relaxation length is proportional to the phonon mean free paths of the corresponding lead material, it is also heavily depends on the material in the opposite lead. A mismatch of acoustic vibrational spectra alters the spectral distribution of the heat flux. Given the strong function of phonon mean free path with respect to the phonon frequency, it results in largely varying relaxation length of non-equilibrium phonons depending on the combination of lead materials.

## V. Acknowledgement

We thank Richard Wilson and Yee Kan Koh for helpful discussions. We acknowledge support from National Science Foundation (Award No. 1943807). This research was also supported in part by the University of Pittsburgh Center for Research Computing, RRID:SCR_022735, through the resources provided. Specifically, this work used the H2P cluster, which is supported by NSF award number OAC-2117681.

**Figures**

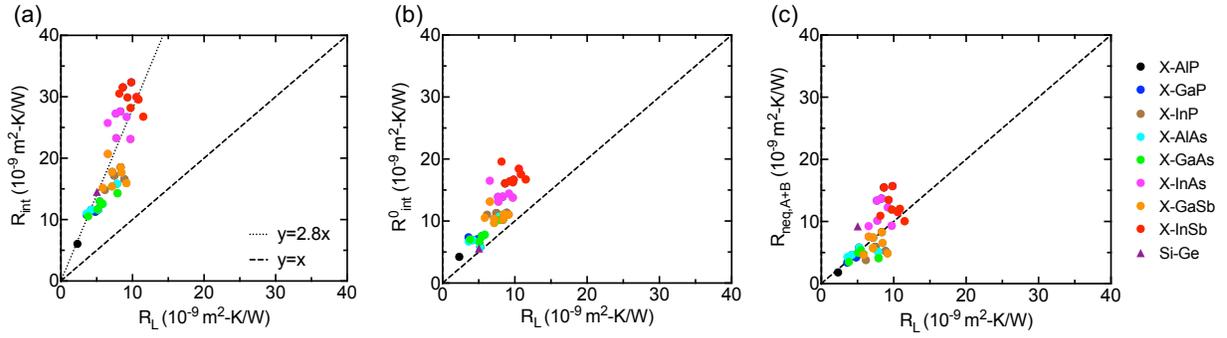

Fig. 1. The thermal resistances from the PBE compared to those from the Landauer formalism: (a) overall interfacial thermal resistance ($R_{int}$), (b) resistance directly caused by interface scattering ($R_{int}^0$), and (c) non-equilibrium thermal resistance in two leads ($R_{neq,A+B}$). The $R_L$ is the resistance from the Landauer formalism. In the legend, the X signifies any III-V compound. Note that heterogeneous interfaces are shown twice; for example, AlP-AlAs interface shown as X-AlP and X-AlAs. For comparison, the Si-Ge interface is represented by a purple triangle[20].

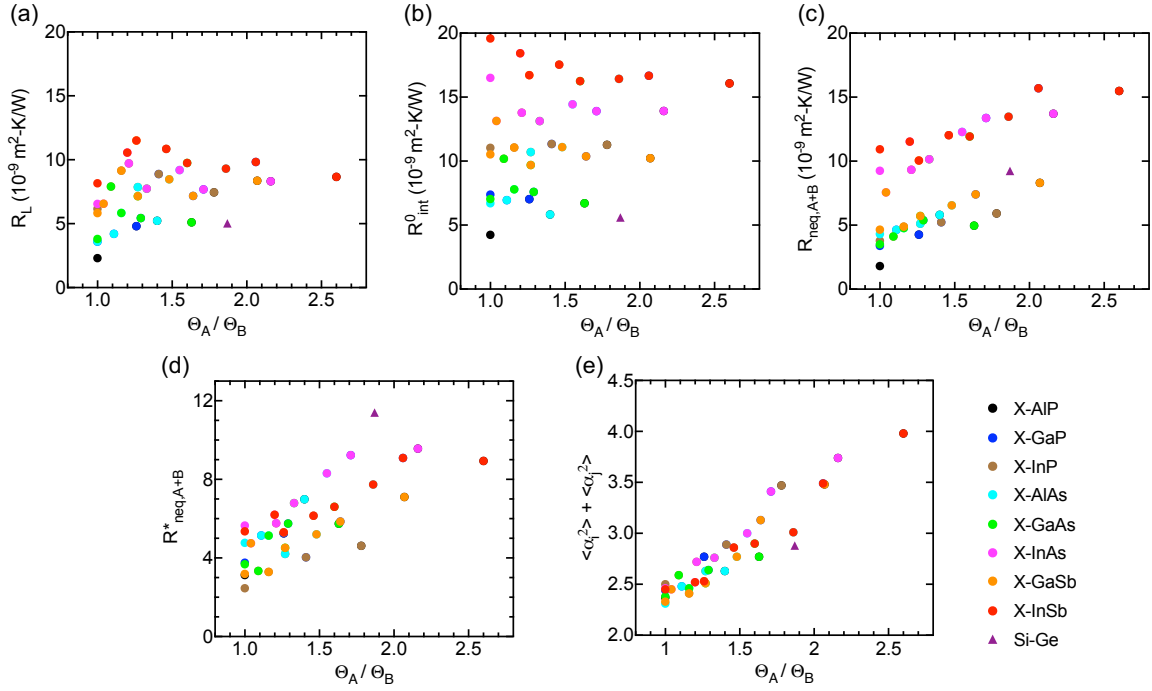

Fig. 2. The resistance versus the Debye temperature ratio $\Theta_A/\Theta_B$ between two lead materials: (a) resistance predicted by Landauer formalism with DMM model $R_L$, (b) resistance directly induced by interface scattering $R_{int}^0$, (c) non-equilibrium resistance $R_{neq,A+B}$, (d) dimensionless non-equilibrium resistance $R^*_{neq,A+B}$ representing $\langle(\alpha_i^2-1)\beta_i\rangle + \langle(\alpha_j^2-1)\beta_j\rangle$, and (e) the degree of non-equilibrium distribution at the interface $\langle\alpha_i^2\rangle + \langle\alpha_j^2\rangle$.

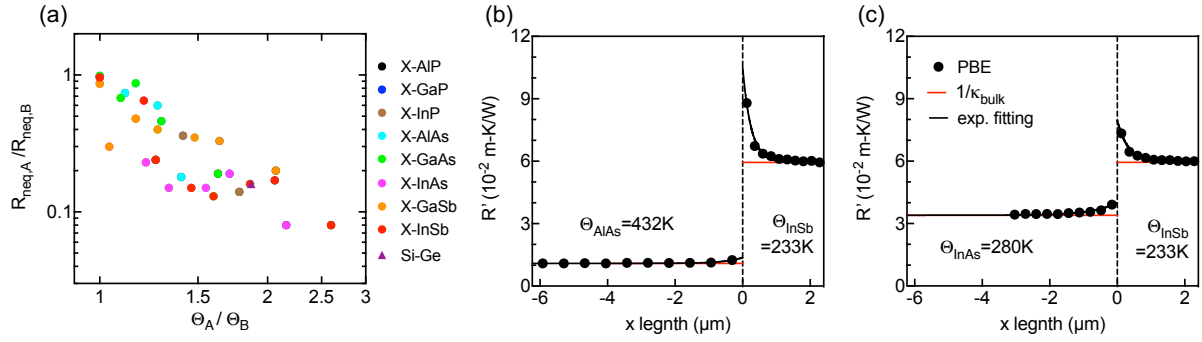

Fig. 3. (a) The non-equilibrium resistance ratio $R_{neq,A}/R_{neq,B}$ versus the Debye temperature ratio $\Theta_A/\Theta_B$. The thermal resistivity $R'$ for (b) AlAs-InSb interface and (c) InAs-InSb interface. In (b-c), the black solid circles represent PBE data, black lines indicate their exponential fits, and the red line denotes the thermal resistivity of bulk materials.

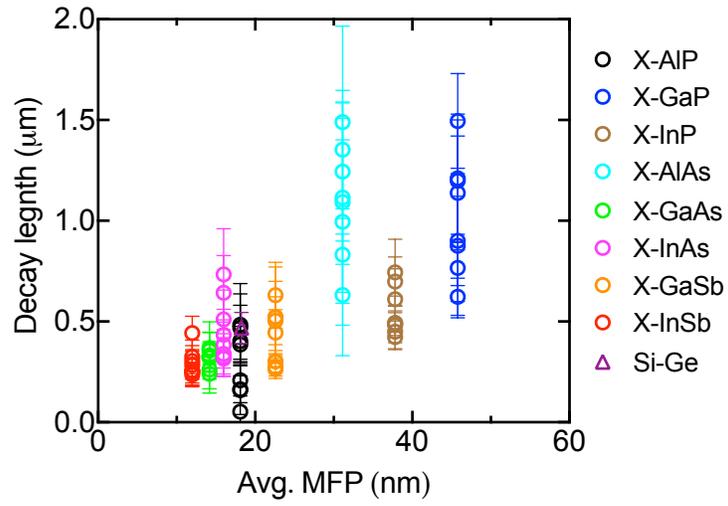

Fig. 4. The relaxation lengths of non-equilibrium resistivity in the right lead compared to the mode-averaged mean free paths of the corresponding lead materials.

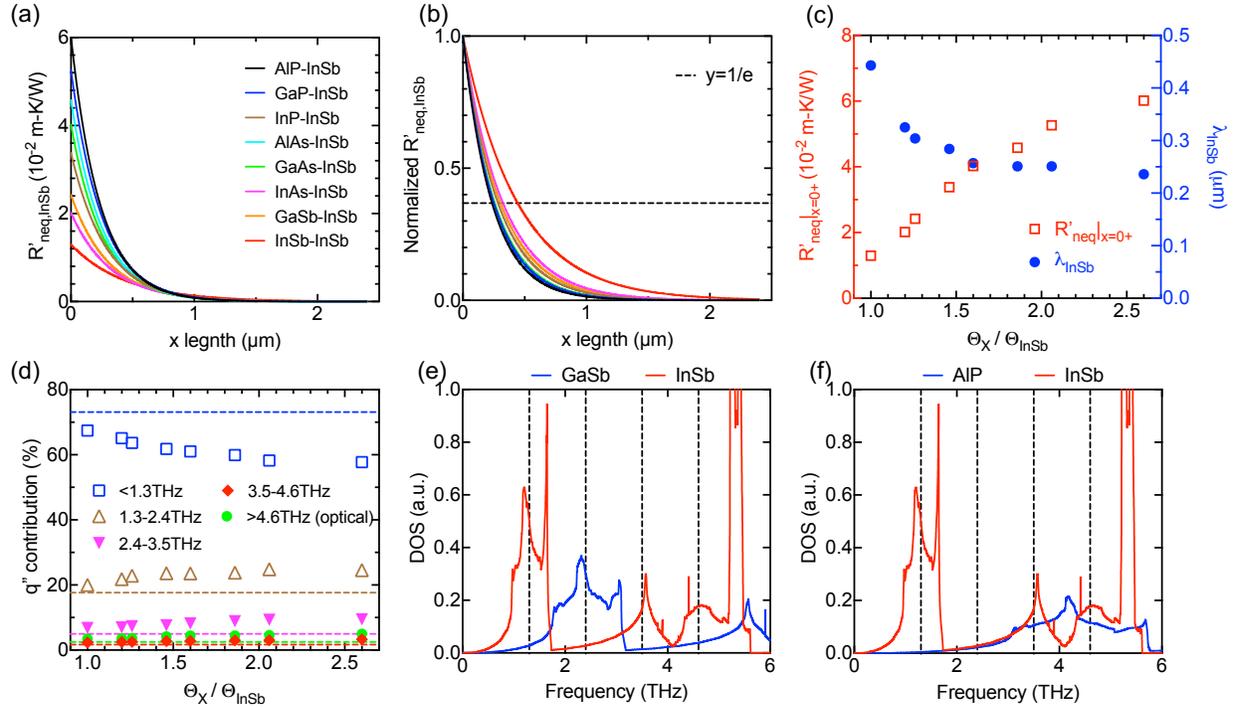

Fig. 5. Effects of lead material X on the non-equilibrium resistivity and its decay in X-InSb interfaces. (a-b) Spatial decay of non-equilibrium resistivity in InSb lead. The horizontal dashed line in (b) represents $y=1/e$ to indicate the relaxation length. (c) The non-equilibrium resistivity at the interface (red open square) and relaxation length (blue solid circles) in InSb with respect to Debye temperature ratio for different X-InSb interfaces. (d) Heat flux contribution from 5 frequency bins sampled at the control volume of InSb closest to the interface. The dashed lines represent the spectral heat flux contribution at the boundary of InSb lead which is far from the interface and exhibits the bulk phonon distribution. (e-f) Overlap of phonon density-of-states in GaSb-InSb ($\Theta_{GaSb}/\Theta_{InSb}=1.26$) and AlP-InSb ($\Theta_{AlP}/\Theta_{InSb}=2.6$) interfaces. The vertical dashed lines are the boundaries of the five frequency bins used in (d).

# Supplemental Materials for

# Non-equilibrium Thermal Resistance of Interfaces Between III-V Compounds

Jinchen Han[1], and Sangyeop Lee[1,2*]

[1] *Department of Mechanical Engineering and Materials Science,*
*University of Pittsburgh, Pittsburgh PA 15261*

[2] *Department of Physics and Astronomy,*
*University of Pittsburgh, Pittsburgh PA 15261*

1. DFT calculation results

We summarize all DFT calculated lattice constants $a$, Debye temperature $\Theta$, averaged mean free path (MFP) $\Lambda_{avg}$, thermal conductivity under relaxation time approximation $\kappa_{RTA}$, and lead lengths $L_{lead}$ in Table I. The averaged MFP of each material is calculated using modal specific heat as

$$\Lambda_{avg} = \sum_i \hbar\omega_i \frac{\partial f_i^0}{\partial T} \Lambda_i \bigg/ \sum_i \hbar\omega_i \frac{\partial f_i^0}{\partial T}. \tag{S1}$$

The thermal conductivity calculation only consider three-phonon scattering under relaxation time approximation. The lead length of any material is $200\Lambda_{avg}$ and same to all interfaces.

TABLE I. Converted conventional lattice constants $a$, Debye temperature $\Theta$, averaged mean free path $\Lambda_{avg}$ (300 K), thermal conductivity under relaxation time approximation $\kappa_{RTA}$ (300K), and lead lengths $L_{lead}$ (300K) for materials considered in this work.

|  | $a$ (Å) | $\Theta$ (K) | $\Lambda_{avg}$ (nm) | $\kappa_{RTA}$ (W/m-K) | $L_{lead}$ (μm) |
|---|---|---|---|---|---|
| AlP | 5.436 | 605.3 | 18.1 | 80.9 | 3.62 |
| GaP | 5.426 | 478.5 | 45.8 | 164 | 9.16 |
| InP | 5.879 | 339.3 | 37.8 | 78.2 | 7.56 |
| AlAs | 5.638 | 432.4 | 31.1 | 91.6 | 6.23 |
| GaAs | 5.627 | 371.5 | 14.2 | 39.7 | 2.84 |
| InAs | 6.060 | 279.7 | 16.0 | 29.4 | 3.20 |
| GaSb | 6.066 | 292.2 | 22.6 | 39.2 | 4.52 |
| InSb | 6.471 | 232.7 | 12.0 | 16.8 | 2.40 |
| Si | 5.468 | 684.0 | 26.9 | 127.5 | 5.40 |
| Ge | 5.782 | 365.5 | 18.3 | 45.8 | 3.70 |